\documentclass[12pt]{iopart}
\usepackage{graphicx} 
\usepackage{float}
\usepackage{geometry}
\usepackage{color}
\usepackage{tikz}
\usetikzlibrary{shapes}
\newcommand{\meanen}{{\left( \langle \epsilon \rangle \right)}}

\newcommand{\sfsix}{SF$_6$ }

\newcommand{\ecrit}{$\left(E/n_0\right)_{c}$}

\begin{document}

\title[Robust approximation rules for critical electric field of dielectric gas mixtures]{Robust approximation rules for critical electric field of dielectric gas mixtures}

\author{N.A. Garland}
\address{Centre for Quantum Dynamics, Griffith University, Nathan 4111, Australia}
\address{School of Environment and Science, Griffith University, Nathan 4111, Australia}
\ead{n.garland@griffith.edu.au}
\author{D.L. Muccignat}
\address{College of Science and Engineering, James Cook University, Townsville 4810, Australia}
\author{G.J. Boyle}
\address{College of Science and Engineering, James Cook University, Townsville 4810, Australia}
\author{R.D. White}
\address{College of Science and Engineering, James Cook University, Townsville 4810, Australia}

\vspace{10pt}

\begin{abstract}
A semi-analytic method for quickly approximating the density-reduced critical electric field for arbitrary mixtures of gases is proposed and validated. Determination of this critical electric field is crucial for designing and testing alternatives to SF$_6$ for insulating high voltage electrical equipment. We outline the theoretical basis of the approximation formula from electron fluid conservation equations, and demonstrate how for binary mixtures the critical electric field can be computed from the transport data of electrons in the pure gases. We demonstrate validity of the method in mixtures of N$_2$ and O$_2$, and SF$_6$ and O$_2$. We conclude with an application of the method to approximate the critical electric field for mixtures of SF$_6$ and HFO1234ze(E), which is a high interest mixture being actively studied for high voltage insulation applications. 
\end{abstract}

\section{Introduction}

Presently, most high voltage electrical infrastructure on Earth relies on sulphur hexafluoride (SF$_6$). This inert gas is utilised in equipment such as circuit breakers, transformers, and transmission infrastructure due its excellent electric insulation and arc-extinguishing capacity~\cite{tian_research_2020}. Identified as a greenhouse gas in the Kyoto protocol~\cite{noauthor_kyoto_2005}, \sfsix has since been labeled by the Intergovernmental Panel on Climate Change as the most potent greenhouse gas with a global warming potential~(GWP) approximately 23,500 times that of carbon dioxide~(CO$_2$), and with an anticipated lifetime of up to 3200 years. 

The bulk of \sfsix emissions come from the electric power industry, often during maintenance of electrical equipment or simply due to aging or faulty storage vessels and electrical equipment. Recently, some success has been reported in legislatively constraining \sfsix emissions~\cite{hu_declining_2023,simmonds_increasing_2020}. However, ultimately the global \sfsix gas content is still increasing year-on-year and growth in newly installed gas-insulated equipment is overpowering industry regulation and financial penalty policies in reducing \sfsix  emissions.~\cite{simmonds_increasing_2020}. 

To date, a number of replacements to \sfsix have been deployed in the power industry, such as dry air, N$_2$, CO$_2$, or various admixtures of natural and proprietary synthetic gases~\cite{tian_research_2020}. Despite some adoption of replacements with lower GWP, the widespread global use of \sfsix prevails in many industrial applications where it has not yet been bettered on an engineering basis for insulation and/or current quenching requirements.  In addition, the logistical and/or financial burdens to replace \sfsix based equipment are often significant.

Driving the effort towards replacing \sfsix are many hours and millions of dollars spent per year in academic and commercial research performing dielectric breakdown experiments~\cite{basu_improved_2023,zeng_breakdown_2022} or swarm experiments~\cite{eguz_synergism_2023,eguz_discussion_2022} to characterise insulating gases. There are many complexities and challenges in adequately performing and interpreting measurements of such experiments in their varying geometries (e.g. needle-plane, parallel plate, coaxial) and varying operating conditions (e.g. low or high pressure/temperature, gas purity, pulsed Townsend or steady state Townsend in the case of swarm experiments). As a result, theory and simulation are vital methods helping to complement experiments.

The power and flexibility of theory and simulation applied to highly unconstrained design problems can enable a quick and affordable scan of very large parameter spaces of material choices or design configurations. This computational sieve approach to rapidly explore, trial, and assess potential scenarios then informs the design of targeted and efficient experiments and solution prototypes. For example, insight through large simulations of breakdown scenarios in often complex geometries \cite{zhang_numerical_2022} or detailed kinetic calculations of electron energy distribution functions (EEDFs) in gas mixtures over density or pressure windows~\cite{flynn_benchmarking_2021} are now capable with advances in computing. That being said, these non-trivial computational tasks can also present their own challenges and barriers to entry, such as advanced computational domain knowledge or access to sufficient computing power or time. 

In addition to brute force computational methods, analytic or simplified semi-analytic methods can be employed to provide very simple and rapid estimations of design parameters. In particular, simple design formulas can be accessed by a wide user base, allowing experiment scenarios to be quickly explored. In Section~\ref{sec:theory} we present a semi-analytic method to rapidly predict the critical reduced electric field, \ecrit, of gas mixture combinations based entirely on calculated electron transport properties of the constituent pure gases only. We first demonstrate the validity of this model in Section~\ref{sec:benchmark} by benchmarking the predictive capability with relatively simple mixtures of well studied gases, such as N$_2$-O$_2$, and SF$_6$-O$_2$. Finally, in Section~\ref{sec:hfo} we demonstrate the ability of the model to correctly predict the positive synergy in \ecrit~for mixtures of \sfsix and HFO-1234ze(E). This molecule~(1E)-1,3,3,3-Tetrafluoroprop-1-ene (C$_3$H$_2$F$_4$), often termed HFO-1234ze(E) or R-1234ze, is a hydrofluoroolefin (HFO) used as a refrigerant and is studied as a potential \sfsix replacement~\cite{chachereau_electron_2016,eguz_discussion_2022}.

\section{Theory}\label{sec:theory}

The macroscopic properties of a gas, like breakdown voltage, are often used to guide design of electrical equipment. Fundamentally, the macroscopic behaviours are underpinned by the microscopic interactions between electrons and gas molecules. Pulsed Townsend~(PT) and steady-state Townsend~(SST) swarm experiments have been used to study electron transport properties in insulating gases, such as drift velocity, diffusion coefficients, and importantly ionisation and attachment rates~\cite{tian_research_2020}. One quantity of immediate relevance to the community is the critical electric field, where the ionisation rate equals the electron attachment rate. Beyond this field the rate of ionisation will exceed electron attachment and so it marks the transition of a gas from being a free electron sink to being a source, precipitating the dielectric breakdown of the medium.

Primary methods of approximating swarm properties in gas mixtures have largely been based on projecting functional dependence onto the reduced electric field, $E/n_0$, which we will abbreviate in equations as $\check{E}=E/n_0$ for brevity. One then approximates quantities of a gas mixture as linear combinations of pure gas quantities evaluated at a common value of $E/n_0$, weighted by gas density fractions of the mixture. For example, for simple gas mixtures, many have employed Blanc's law~\cite{blanc_recherches_1908} to approximate drift velocity
\begin{eqnarray}
\frac{1}{W_{\mathrm{m}}(\check{E})} =  \frac{x}{W_A(\check{E})} + \frac{(1-x)}{W_B(\check{E})},
\label{blanc-law} 
\end{eqnarray}
where $W_{\mathrm{m}}$ is the electron drift velocity in the mixture, $W_A$ and $W_B$ are drift velocities in gas A and gas B, and 
\begin{eqnarray}
x=\frac{n_A}{n_A + n_B}, \label{eq:x-frac} 
\end{eqnarray}
where $n_A$ and $n_B$ are present gas densities of constituent gases A and B evaluated at a common value of $E/n_0$. 

In a similar fashion, a so-called Wieland approximation has been used to approximate ionisation, $k_{\mathrm{i,m}}$, or attachment, $k_{\mathrm{a,m}}$, rates of gas mixtures~\cite{wieland_gasdurchschlagmechanismen_1973}
\begin{eqnarray}
k_{\mathrm{m}}(\check{E}) = x k_{\mathrm{A}}(\check{E}) + (1-x)k_{\mathrm{B}}(\check{E}),
\label{eq:wieland} 
\end{eqnarray}
where $k_{\mathrm{m}}$ is a rate in the mixture and $k_{\mathrm{A}}$ and $k_{\mathrm{B}}$ are the rates in pure gases A and B evaluated at a common value of $E/n_0$.

Implicit in its derivation, Blanc's Law~\cite{blanc_recherches_1908} does not inherently include inelastic collisions. Further assumptions inherent in the Blanc and Wieland approximations that limit their validity are that (i) electron impact cross sections can be added in simple linear combinations at a given value of $E/n_0$, and (ii) the steady state EEDF is the same for each component gas, and the combination mixture, at a given value of $E/n_0$. This can be a reasonably good approximation for some certain conditions, however in general this assumption of the EEDF can be overly restrictive.

Previous studies have highlighted shortcomings in the Wieland approximation~\cite{maric_parametrization_2005,van_brunt_common_1987}, with one notable limitation being an inability to predict positive synergism in gas mixtures - where a higher critical electric field of the mixture, above that of either constituent gases, is obtained. To remedy shortcomings, extensions on this style of approximation, allowing for a non-linear perturbation around the basic linear combination, have been explored~\cite{eguz_synergism_2023,eguz_discussion_2022}. Indeed, van Brunt~\cite{van_brunt_common_1987} proposed a novel alternative parameterisation of swarm properties via consideration of electron temperatures, similar to what is presented in this current study. However, while this parameterisation identifies the crucial role of mean electron energy, it relies on multiple free parameters that require specification for each gas being considered.

To better account for the dependence of energy transfer on inelastic collisions as $E/n_0$ increases an alternative approach to Blanc's law based on electron mean energy was developed~\cite{chiflikian_analog_1995}. This alternative approach for evaluating collision and transport data of electrons in gas mixtures assumes equivalence of a shared common electron mean energy (CME), $\langle\varepsilon\rangle$, instead of a common reduced electric field, $E/n_0$~\cite{chiflikian_analog_1995}. This assumption therefore allows for arbitrary EEDFs in either constituent gas or the mixture itself, with the restriction that the integrated moment of $mv^2/2$ over each EEDF must be equivalent.
The CME approach has been well adopted by many areas of the community, however use of Blanc's Law still persists in some studies. 

Of particular note to this study is the prior demonstration~\cite{maric_parametrization_2005} of using the CME approximation for estimation of non-conservative collision rates, $\nu^*$, in mixtures of molecular gases
\begin{eqnarray}
\nu^*_{\mathrm{m}}(\langle \epsilon \rangle) = x\nu^*_{\mathrm{A}}(\langle \epsilon \rangle) + (1-x)\nu^*_{\mathrm{B}}(\langle \epsilon \rangle), \label{eq:nu-star-sum} 
\end{eqnarray}
this assumption was shown to be a much better method of projecting complex, non-linear electron kinetic effects onto a single macroscopic swarm variable. In arriving at the semi-analytic model presented in this study, we will also adopt the approximation for estimation of non-conservative collision rates in equation (\ref{eq:nu-star-sum}), and leverage the CME approach to determine $E/n_0$ in the mixture conditions; this will enable estimation of the critical reduced electric field. 

As a foundation for our approach, we refer to the momentum and energy conservation equations for electron transport, that can be derived by taking integral moments of the Boltzmann equation over trial functions for electron momentum, $m\textbf{v}$, and kinetic energy, $mv^2/2$~\cite{garland_unified_2017,simonovic_electron_2019,boyle_boltzmanns_2023,garland_electron_2018}. As per the CME approximation idea~\cite{chiflikian_analog_1995}, it has been previously demonstrated~\cite{chiflikian_analog_1995,garland_approximating_2018} that in the steady state limit the electron momentum balance equations of both pure gases A and B, and an arbitrary mixture of the two, yield the following expression for the electron drift velocity in the mixture
\begin{eqnarray}
\frac{1}{W_{\mathrm{m}}\meanen} = x \frac{\check E_{\mathrm{A}}\meanen}{\check E_{\mathrm{m}}\meanen}\frac{1}{W_{\mathrm{A}}\meanen} + (1-x) \frac{\check E_{\mathrm{B}}\meanen}{\check E_{\mathrm{m}}\meanen}\frac{1}{W_{\mathrm{B}}\meanen}.
\label{eq:mom-bal-drift} 
\end{eqnarray}

While from the steady state energy balance equations of both pure gases A and B, and an arbitrary mixture of the two, a second expression for the electron drift velocity in the mixture can be found
\begin{eqnarray}
W_{\mathrm{m}}\meanen = x \frac{\check E_{\mathrm{A}}\meanen}{\check E_{\mathrm{m}}\meanen}W_{\mathrm{A}}\meanen + (1-x) \frac{\check E_{\mathrm{B}}\meanen}{\check E_{\mathrm{m}}\meanen}W_{\mathrm{B}}\meanen.
\label{eq:en-bal-drift} 
\end{eqnarray}

With (\ref{eq:mom-bal-drift}) and (\ref{eq:en-bal-drift}) both explicitly referring to the value of $E/n_0$ in the mixture, we can combine (\ref{eq:mom-bal-drift}) and (\ref{eq:en-bal-drift}) to obtain either an expression for the drift velocity if one desires 
\begin{eqnarray}
W^2_{\mathrm{m}}\meanen = \frac{x \check E_{\mathrm{A}} W_{\mathrm{A}} + (1-x)\check E_{\mathrm{B}} W_{\mathrm{B}}  }{x\check E_{\mathrm{A}} W_{\mathrm{B}} + (1-x)\check E_{\mathrm{B}} W_{\mathrm{A}}}W_{\mathrm{A}} W_{\mathrm{B}} ,
\label{eq:drift_final} 
\end{eqnarray}
or an expression for the reduced electric field in the mixture
\begin{eqnarray}
\check E^2_{\mathrm{m}}\meanen = \frac{\left(x \check E_{\mathrm{A}} W_{\mathrm{A}} + (1-x)\check E_{\mathrm{B}} W_{\mathrm{B}} \right) \left(x\check E_{\mathrm{A}} W_{\mathrm{B}} + (1-x)\check E_{\mathrm{B}} W_{\mathrm{A}}\right)}{W_{\mathrm{A}} W_{\mathrm{B}}} ,
\label{eq:efield_final} 
\end{eqnarray}
where $\check E_{\mathrm{A}}=\check E_{\mathrm{A}}\meanen$, $\check E_{\mathrm{B}}=\check E_{\mathrm{B}}\meanen$, $W_{\mathrm{A}}=W_{\mathrm{A}}\meanen$, and $W_{\mathrm{B}}=W_{\mathrm{B}}\meanen$ are all functions of the common mean electron energy being considered, $\langle \epsilon \rangle$.

\subsection{Approximation rule for critical reduced electric field}
By using the earlier approximation for a mixture ionisation or attachment collision frequency in equation (\ref{eq:nu-star-sum}), and the expression in equation (\ref{eq:efield_final}) for equivalent reduced electric field in the mixture, we can now state the procedure to extract a value of critical electrical field as:
\begin{enumerate}
    \item Using a platform of choice~(e.g. Boltzmann equation solver, Monte Carlo simulation) compute steady state electron swarm parameters for pure gases A and B. 
    \item Extract necessary variables: reduced electric field ($E/n_0$), drift velocity ($W$), mean energy ($\langle \epsilon \rangle$), ionisation coefficient ($k_{\mathrm{i}}$), attachment coefficient ($k_{\mathrm{a}}$).
    \item Using the available swarm data and desired density fraction given by (\ref{eq:x-frac}), evaluate ionisation and attachment coefficients of a mixture as a function of mean energy
    \begin{eqnarray}
    k_{\mathrm{i,m}}\meanen = x k_{\mathrm{i,A}}\meanen + (1-x)k_{\mathrm{i,B}}\meanen,
    \label{eq:k-ion} 
    \end{eqnarray}
    \begin{eqnarray}
    k_{\mathrm{a,m}}\meanen = x k_{\mathrm{a,A}}\meanen + (1-x)k_{\mathrm{a,B}}\meanen.
    \label{eq:k-att} 
    \end{eqnarray}
    \item Using the available data, evaluate (\ref{eq:efield_final}) to obtain the reduced electric field, $\check E_{\mathrm{m}}\meanen$, in the mixture as a function of mean energy.
    \item Evaluate the net effective ionisation coefficient
    \begin{eqnarray}
    k_{\mathrm{eff,m}}\meanen = k_{\mathrm{i,m}}\meanen - k_{\mathrm{a,m}}\meanen,
    \label{eq:k-eff} 
    \end{eqnarray}
    which can now be expressed, or plotted for convenience, using $\check E_{\mathrm{m}}\meanen$ as a function $k_{\mathrm{eff,m}}\left(E/n_0\right)$ to determine the zero-crossing and thus the critical reduced electric field of the gas mixture.    
\end{enumerate}

So, with zero knowledge of the electron transport in the mixture, we have detailed an approximation of the critical reduced 
electric field of a gas mixture. This uses only the electron swarm data computed for pure gases, which is a standard exercise being made evermore accessible by multiple freely available software for download, or even codes evaluated online via a web browser~\cite{hagelaar_solving_2005,rabie_methes_2016,stephens_multi-term_2018,park2023thunderboltz}. To validate the use of this presented method, we now present benchmarking results of approximating the critical reduced electric field for mixtures of relatively well-studied molecular gases.

\section{Benchmarking and model validation}\label{sec:benchmark}
To validate the proposed method to rapidly approximate the critical reduced electric field, we compare to the calculated critical reduced electric field determined for various mixture combinations. The calculated value is determined by first solving the Boltzmann kinetic equation for the EEDF in a mixture gas using a well-benchmarked multi-term solution technique and software framework~\cite{boyle_boltzmanns_2023,white_electron_2018,boyle_multi-term_2017}. One can then compute the effective net electron creation rate, $k_{\mathrm{eff,m}}$, as the difference of the integrated ionisation and attachment rates $k_{\mathrm{i,m}} - k_{\mathrm{a,m}}$. The critical reduced electrical field is naturally determined when $k_{\mathrm{eff,m}}=0$.

We first present results for a simplified case of N$_2$ and O$_2$ mixtures, since only the O$_2$ target will contribute to electron attachment processes. We then follow with benchmarking of SF$_6$ and O$_2$ mixtures, where the addition of SF$_6$ provides a large attachment cross-section to rigorously test the proposed approximation. For the calculations detailed in this study electron scattering cross sections for SF$_6$, O$_2$, and N$_2$ were obtained from the Biagi database\footnote[1]{Biagi-v7.1 database, www.lxcat.net, retrieved on November 30, 2023} of the LXCat online database~\cite{pitchfordlxcat}. 

\subsection{Benchmark: N$_2$ and O$_2$ mixtures }
Mixtures of molecular nitrogen and oxygen are often studied as a surrogate for air. As a result, there has been a wealth of scientific data and understanding produced for electron scattering cross-sections and subsequent transport in pure N$_2$, O$_2$, and indeed mixtures of the two. Therefore, it is a good candidate system for benchmarking and validation of theory and methods. 

\begin{figure}[H]
    \centering
    \includegraphics[width=0.9\linewidth]{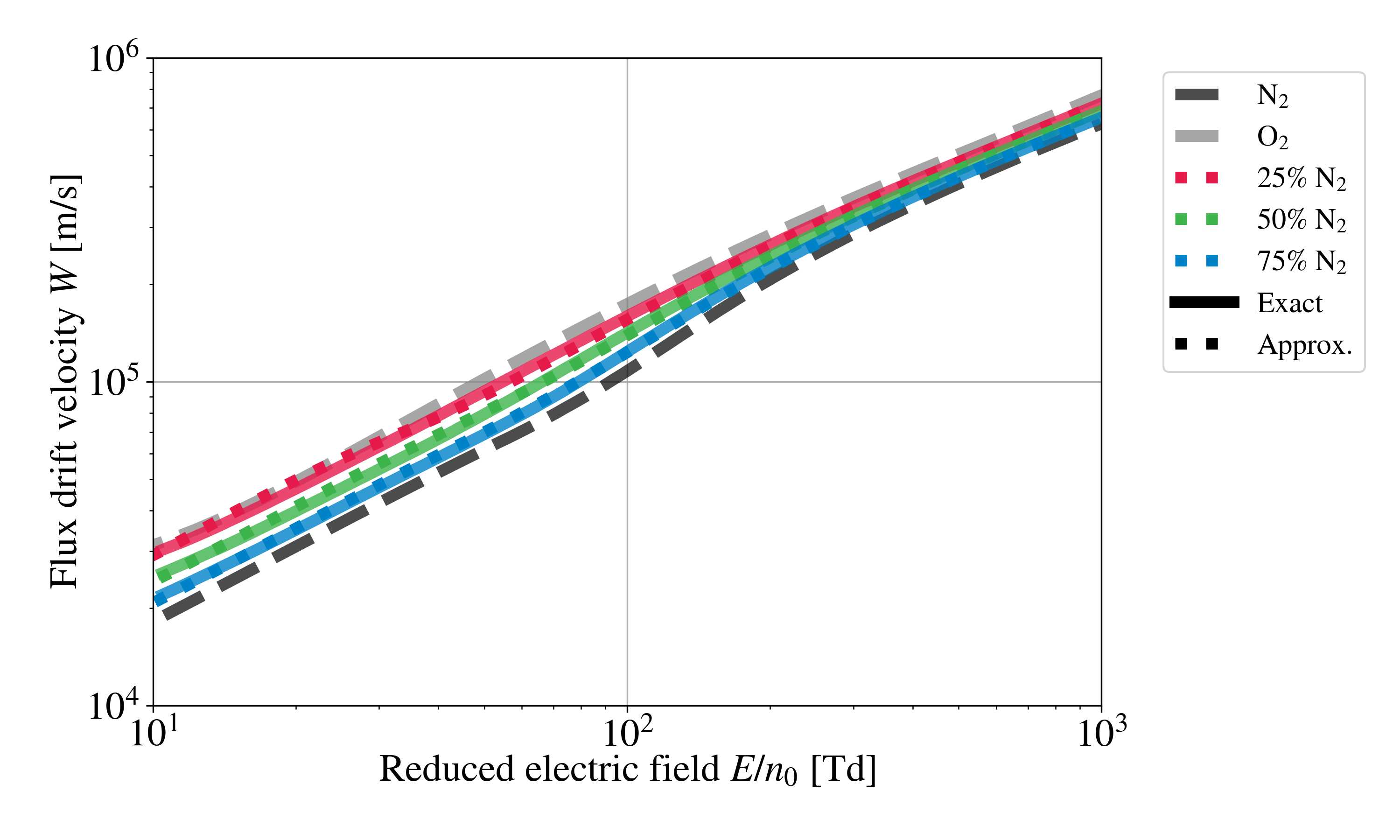}\linebreak
    \includegraphics[width=0.9\linewidth]{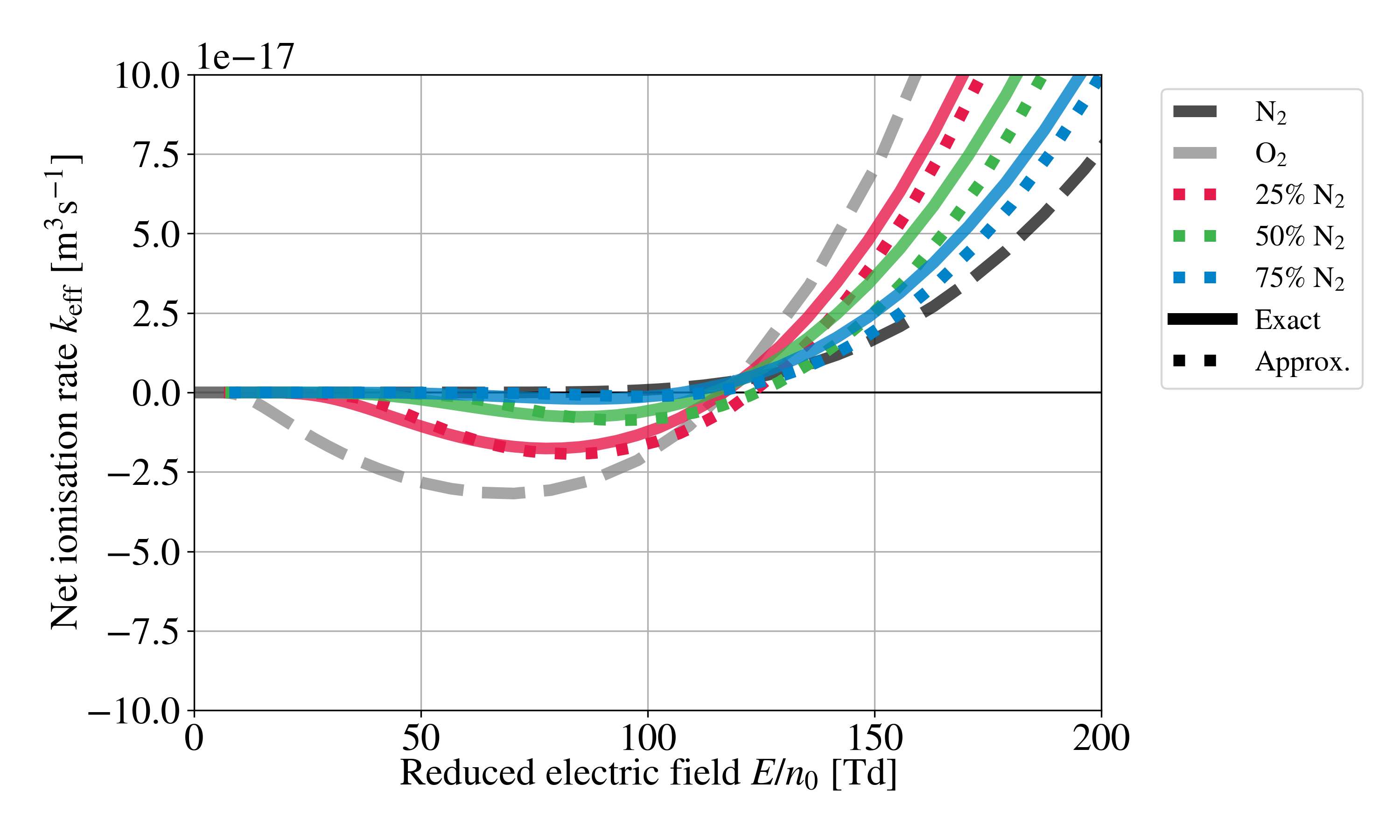}
    \caption{Comparison of (top) flux drift velocity, $W$, and (bottom) net effective ionisation rate, $k_{\mathrm{eff}}$, for various mixtures of N$_2$ and O$_2$ as a function of $E/n_0$. Exact values computed via multi-term solution of the EEDF from the Boltzmann equation, approximations determined from procedure proposed in this study. Horizontal axis in $k_{\mathrm{eff}}$ plot drawn in thicker line to highlight where zero-crossing of curves denotes critical reduced electric field.}
    \label{fig:n2o2drift}
\end{figure}

For validation in the case of N$_2$ and O$_2$ mixtures, in Figure~\ref{fig:n2o2drift} we present the results of approximating both the drift velocity, $W$, and net effective ionisation rate, $k_{\mathrm{eff}}$, as a function of $E/n_0$; for clarity of plot presentation we show comparisons for mixture fractions at 25\%, 50\%, and 75\% ratios.

The present study is chiefly concerned with the net effective ionisation rate, however since drift velocity is also a common transport quantity used in assessment of swarm analysis we include it in Figure~\ref{fig:n2o2drift} as an additional demonstration of the applicability of the approximation method presented in this work. For both transport quantities the approximation produces excellent qualitative agreement compared to the exact calculations. As a quantitative measure of the critical reduced electric field accuracy, Table~\ref{tab:n2o2} shows that, at worst, the agreement is about 5-6\%, which was representative of the general case for other mixture ratios not shown in Figure~\ref{fig:n2o2drift}.

\begin{table}[H]
    \centering
    \caption{Errors in critical reduced electric field approximation for various mixtures of N$_2$ and O$_2$.}
    \label{tab:n2o2}
    \begin{tabular}{cccc}
    \cline{2-3}
    \multicolumn{1}{c|}{}            & \multicolumn{2}{c|}{Critical $E/n_0$ {[}Td{]}}                                   &                                              \\ \cline{2-4} 
    \multicolumn{1}{c|}{}            & \multicolumn{1}{c|}{\textbf{Actual}} & \multicolumn{1}{c|}{\textbf{Approximate}} & \multicolumn{1}{c|}{\textbf{Relative error}} \\ \hline
    \multicolumn{1}{|c|}{25\% N$_2$} & \multicolumn{1}{c|}{117}             & \multicolumn{1}{c|}{122}                  & \multicolumn{1}{c|}{4.3\%}                   \\ \hline
    \multicolumn{1}{|c|}{50\% N$_2$} & \multicolumn{1}{c|}{116}             & \multicolumn{1}{c|}{123}                  & \multicolumn{1}{c|}{6.0\%}                   \\ \hline
    \multicolumn{1}{|c|}{75\% N$_2$} & \multicolumn{1}{c|}{109}             & \multicolumn{1}{c|}{112}                  & \multicolumn{1}{c|}{2.8\%}                   \\ \hline
                                     &                                      &                                           &                                             
    \end{tabular}
\end{table}

\subsection{Benchmark: SF$_6$ and O$_2$ mixtures }
Mixtures of sulphur hexafluoride, SF$_6$, and molecular oxygen, O$_2$, have been well studied in the semiconductor etching industry. As it was for our previous benchmark case, choosing a well studied pair of molecules is a wise choice for benchmarking and validation of theory and methods. In particular, validation in the presence of a highly attaching gas like SF$_6$ accompanied by another attaching gas is worth demonstrating before trialing the methods presented in this study on more complex molecules such as HFO-1234ze(E). 

\begin{figure}[H]
    \centering
    \includegraphics[width=0.9\linewidth]{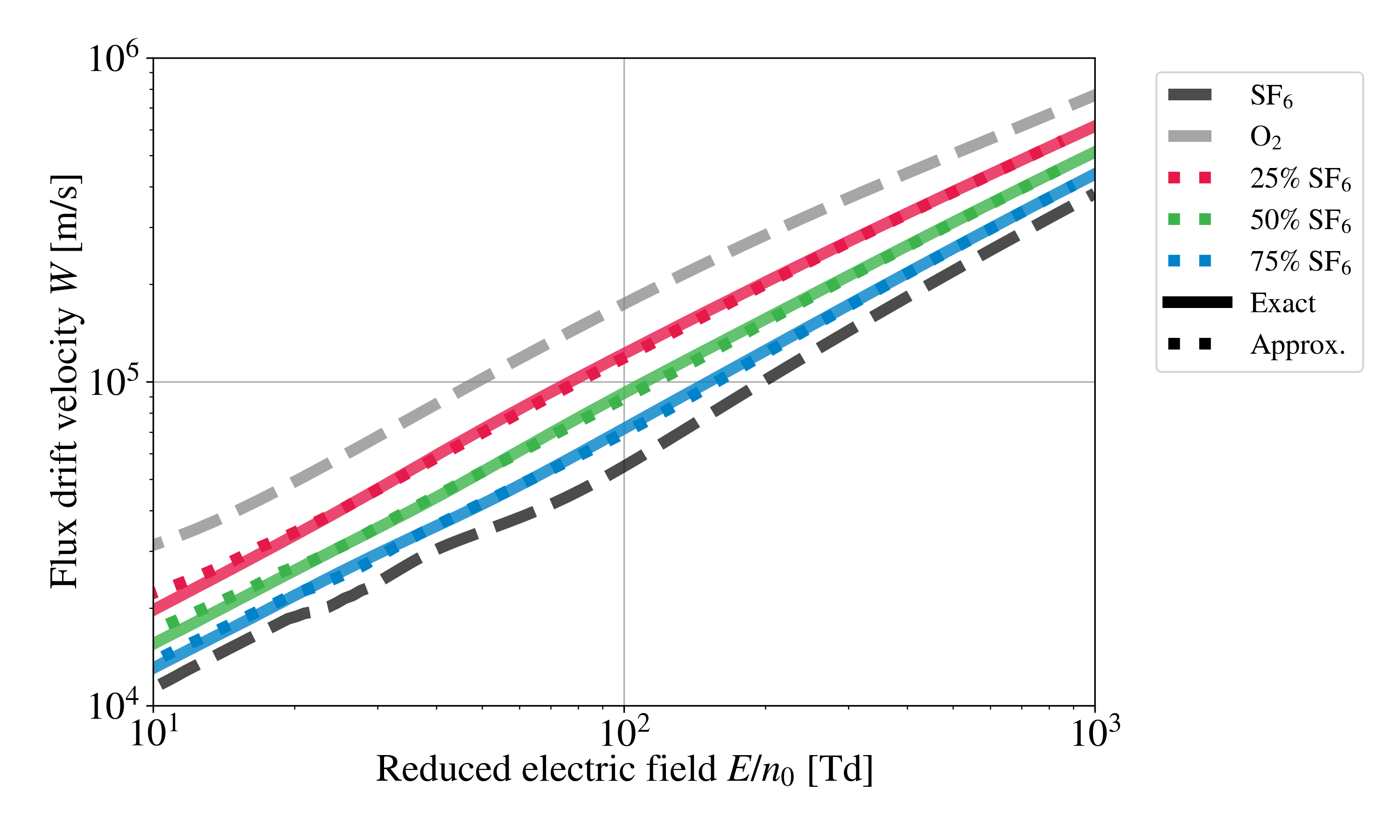}\linebreak
    \includegraphics[width=0.9\linewidth]{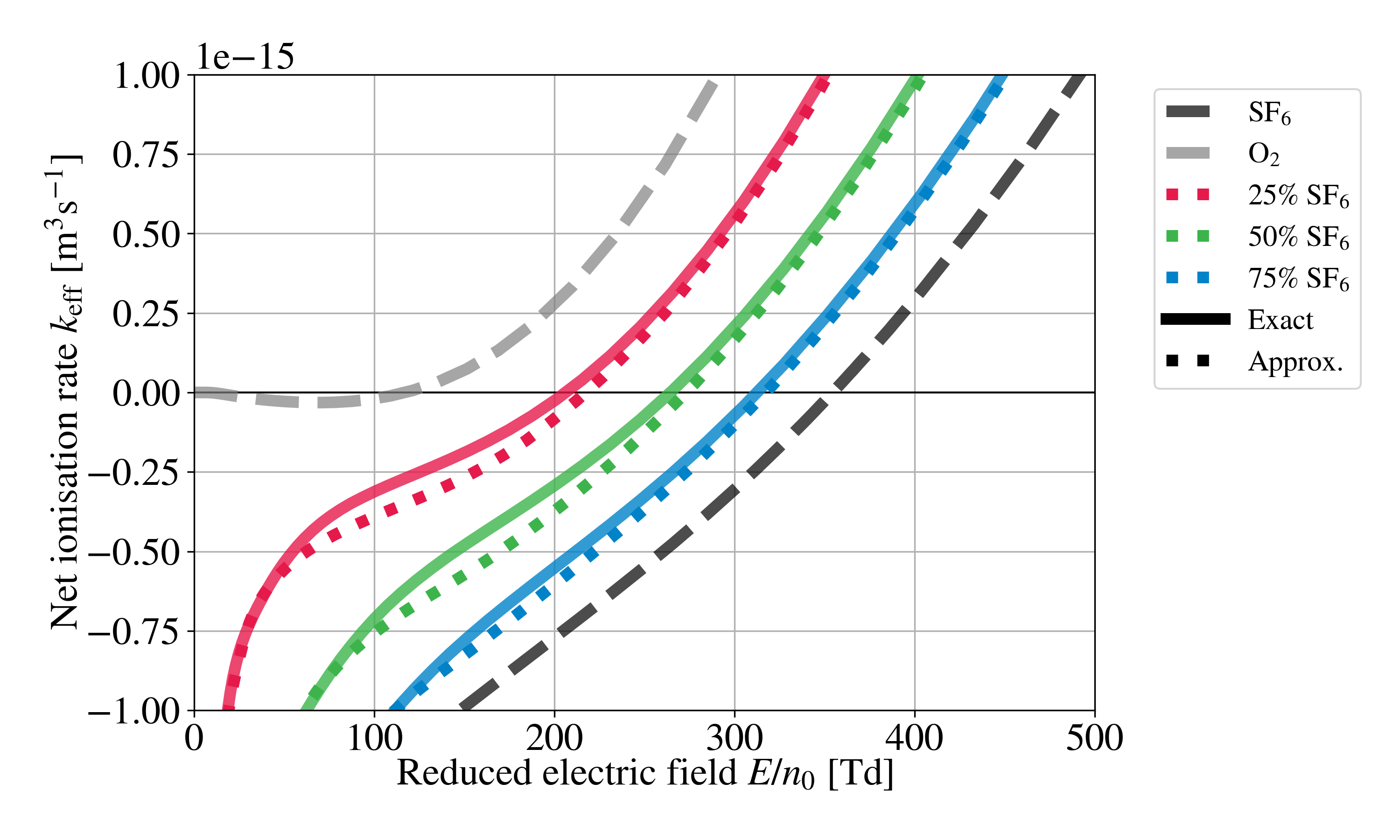}
    \caption{Comparison of (top) flux drift velocity, $W$, and (bottom) net effective ionisation rate, $k_{\mathrm{eff}}$, for various mixtures of SF$_6$ and O$_2$ as a function of $E/n_0$. Exact values computed via multi-term solution of the EEDF from the Boltzmann equation, approximations determined from procedure proposed in this study. Horizontal axis in $k_{\mathrm{eff}}$ plot drawn in thicker line to highlight where zero-crossing of curves denotes critical reduced electric field.}
    \label{fig:sf6o2drift}
\end{figure}

For validation in the case of SF$_6$ and O$_2$ mixtures, in Figure~\ref{fig:sf6o2drift} we present the results of approximating both the drift velocity, $W$, and net effective ionisation rate, $k_{\mathrm{eff}}$, as a function of $E/n_0$; for clarity of presentation we show comparisons for mixture fractions at 25\%, 50\%, and 75\% ratios.

Once again, Figure~\ref{fig:sf6o2drift} demonstrates excellent qualitative agreement produced by the approximation, for both drift velocity and net creation rate. In these cases the transition from pure SF$_6$ to O$_2$ extremes is somewhat monotonic. In terms of the critical reduced electric field values, Table~\ref{tab:sf6o2} values of relative error summarise that there is generally improved accuracy compared with the prior N$_2$/O$_2$ benchmark case. 

\begin{table}[H]
    \centering
    \caption{Errors in critical reduced electric field approximation for various mixtures of SF$_6$ and O$_2$.}
    \label{tab:sf6o2}
    \begin{tabular}{cccc}
    \cline{2-3}
    \multicolumn{1}{c|}{}             & \multicolumn{2}{c|}{Critical $E/n_0$ {[}Td{]}}                                   &                                              \\ \cline{2-4} 
    \multicolumn{1}{c|}{}             & \multicolumn{1}{c|}{\textbf{Actual}} & \multicolumn{1}{c|}{\textbf{Approximate}} & \multicolumn{1}{c|}{\textbf{Relative error}} \\ \hline
    \multicolumn{1}{|c|}{25\% SF$_6$} & \multicolumn{1}{c|}{206}             & \multicolumn{1}{c|}{217}                  & \multicolumn{1}{c|}{5.3\%}                   \\ \hline
    \multicolumn{1}{|c|}{50\% SF$_6$} & \multicolumn{1}{c|}{266}             & \multicolumn{1}{c|}{273}                  & \multicolumn{1}{c|}{3.0\%}                   \\ \hline
    \multicolumn{1}{|c|}{75\% SF$_6$} & \multicolumn{1}{c|}{313}             & \multicolumn{1}{c|}{317}                  & \multicolumn{1}{c|}{1.3\%}                   \\ \hline
                                      &                                      &                                           &                                             
    \end{tabular}
\end{table}

\begin{figure}[H]
    \centering
    \includegraphics[width=1.1\linewidth]{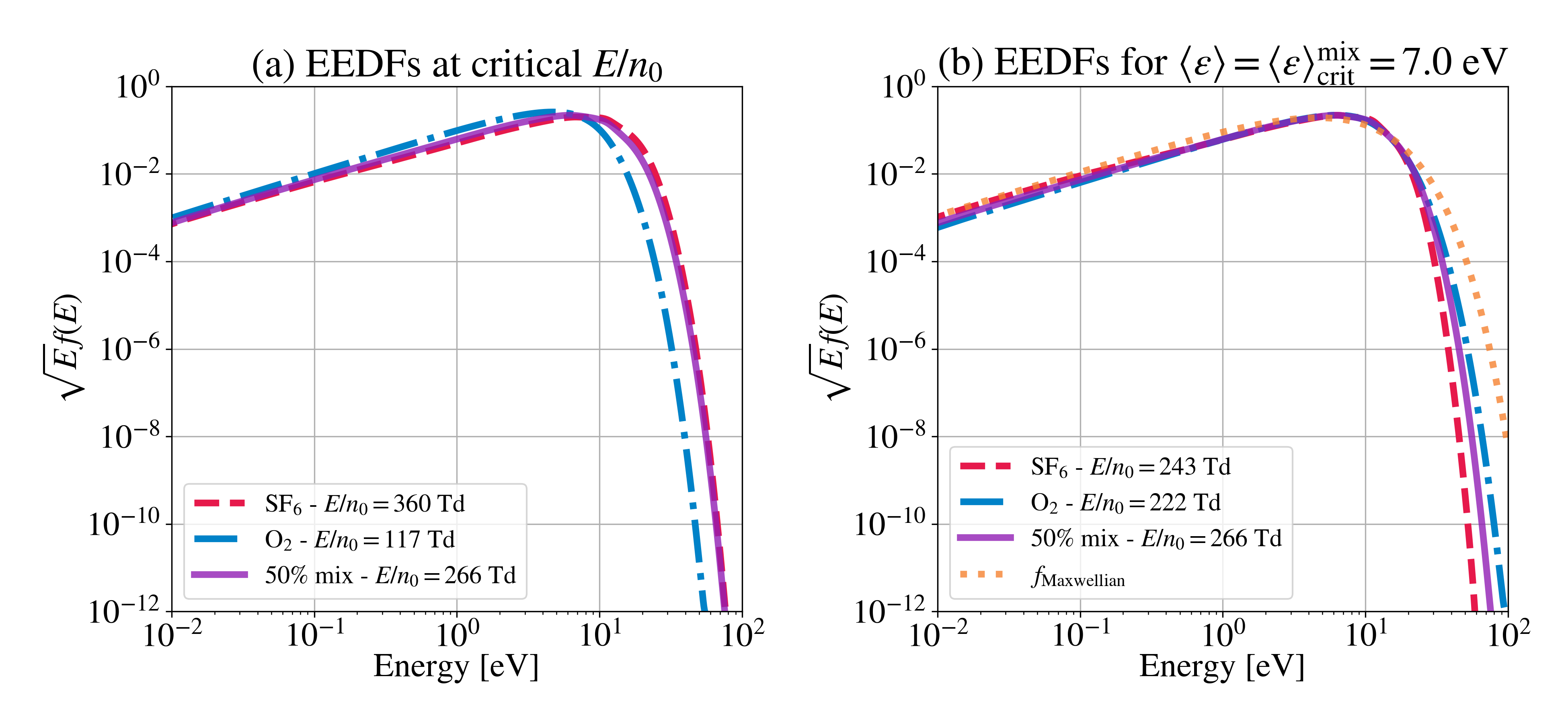}
    \caption{Comparison of EEDFs for SF$_6$, O$_2$, and a mixture of 50\% SF$_6$ and 50\% O$_2$. (Left) EEDFs at each critical $E/n_0$ value for the three background gas cases. (Right) EEDFs for each background gas case associated with a mean electron energy $\langle \epsilon \rangle=7.0$ eV corresponding to the mixture EEDF at the critical field threshold.}
    \label{fig:eedf1}
\end{figure}

As an additional point of interest, Figure~\ref{fig:eedf1} shows the EEDFs at (a) the critical reduced electric field point for pure SF$_6$, O$_2$, and a mixture of 50\% SF$_6$ and 50\% O$_2$. Here, subfigure (a) shows that for the wide range of critical $E/n_0$ values there is some variation in EEDFs. The mixture EEDF generally resembles that for SF$_6$, with some deviation around the distribution peak. Notably the EEDF in O$_2$ peaks at a lower energy. Subfigure (b) shows the same EEDF for the mixture as in (a), however for pure SF$_6$ and O$_2$ the EEDFs are shown corresponding to the same electron mean energy value at the mixture's critical $E/n_0$, $\langle \epsilon \rangle=7.0$ eV. This demonstrates that for the bulk of the distribution population, the EEDFs are indeed quite similar, particularly between approximately 0.1-20 eV. This demonstration highlights that abstracting the dynamics of the bulk population of EEDFs onto the common mean energy $\langle \epsilon \rangle$, invoked in the derivation of our approximation method, is indeed a reasonable approach.

When discussing the nature of EEDFs it is often instructive to compare against a Maxwellian distribution often used by the community. For this purpose, Figure~\ref{fig:eedf1}(b) also allows comparison against a Maxwellian EEDF of the same $\langle \epsilon \rangle$ as the calculated distributions. This highlights that while the bulk of these computed EEDFs reasonably follow the Maxwellian shape about the peak, the sharp roll-off to a depleted tail of higher energy electrons is not sufficiently captured by a Maxwellian.

\section{Application to dielectric insulator design} \label{sec:hfo}
In the search for alternatives to sulphur hexafluoride, SF$_6$, industry and scientific communities are considering a wide range of options. One such option that has been heavily studied~\cite{chachereau_characterization_2017,chachereau_electron_2016,basu_improved_2023,li_partial_2023} is (1E)-1,3,3,3-Tetrafluoroprop-1-ene (C$_3$H$_2$F$_4$), often abbreviated to HFO-1234ze(E). This gas is one of an emerging generation of hydrofluoroolefins (HFOs) originally targeted for use as refrigerants in place of traditional hydrofluorocarbons. The industry use of HFOs is being spurred by their often very low ozone depeletion and global warming potentials.  

Mixtures of sulphur hexafluoride, SF$_6$, and HFO-1234ze(E), are also being studied to maximise insulating and arc quenching properties~\cite{eguz_discussion_2022,hosl_positive_2020} as part of the large design and development research efforts to reduce SF$_6$ use in the high voltage electricity industry. For the HFO-1234ze(E) electron scattering cross-section data, a recently obtained cross-section set from a forthcoming study~\cite{muccignat2023iterative} applying neural network based swarm inversion techniques~\cite{stokes_determining_2020,stokes_self-consistent_2020} is employed. 

\begin{figure}[H]
    \centering
    \includegraphics[width=0.9\linewidth]{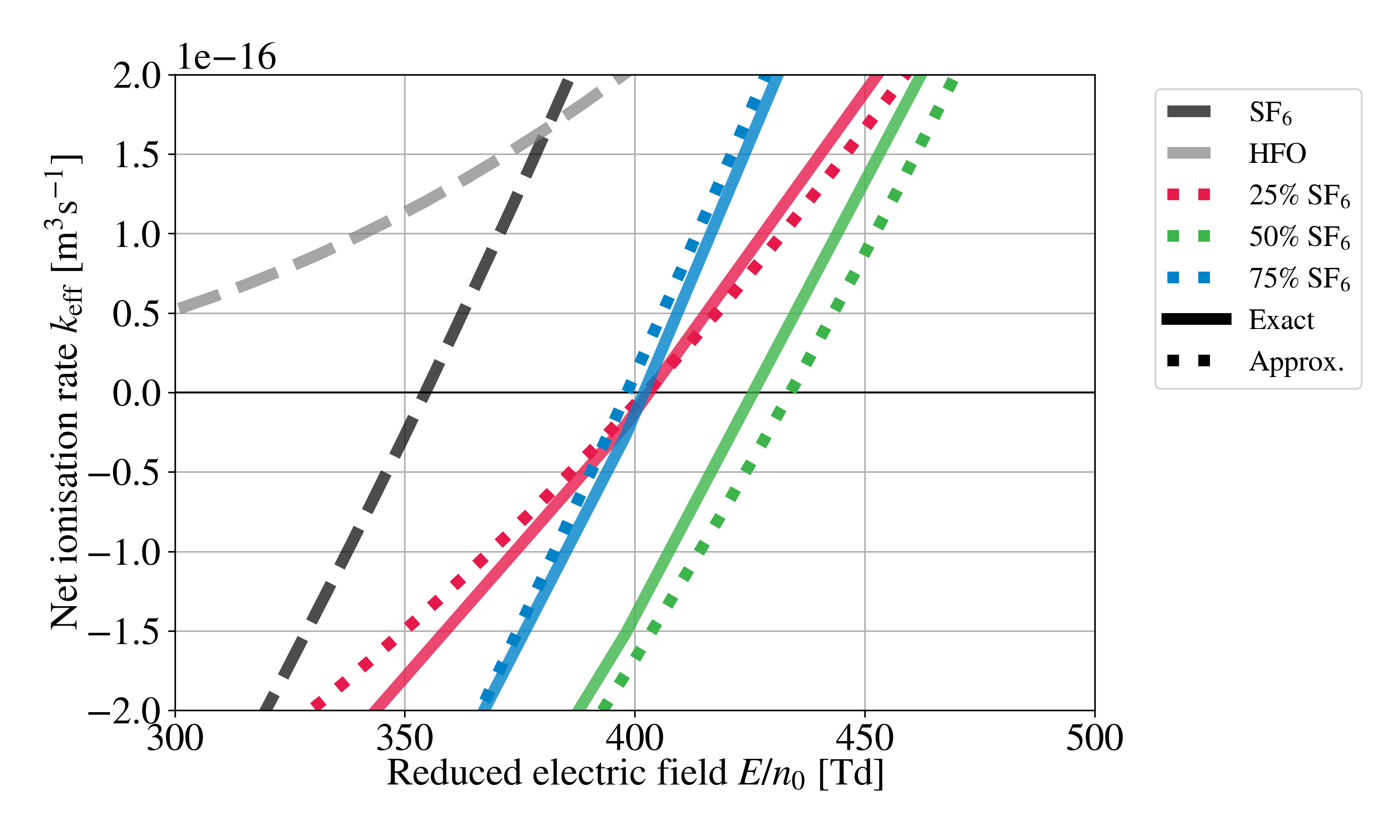}
    \caption{Comparison of net effective ionisation rate, $k_{\mathrm{eff}}$, for various mixtures of SF$_6$ and HFO-1234ze(E) as a function of $E/n_0$. Exact values computed via multi-term solution of the EEDF from the Boltzmann equation, approximations determined from procedure proposed in this study. Horizontal axis in $k_{\mathrm{eff}}$ plot drawn in thicker line to highlight where zero-crossing of curves denotes critical reduced electric field.}
    \label{fig:rate_field}
\end{figure}

Here in Figure~\ref{fig:rate_field} we focus on probing mixtures of SF$_6$ and HFO-1234ze(E) for approximating the net effective ionisation rate, $k_{\mathrm{eff}}$, as a function of $E/n_0$. For clarity of presentation we show comparisons for mixture fractions of 25\%, 50\%, and 75\%.

\begin{figure}[H]
    \centering
    \includegraphics[width=1\linewidth]{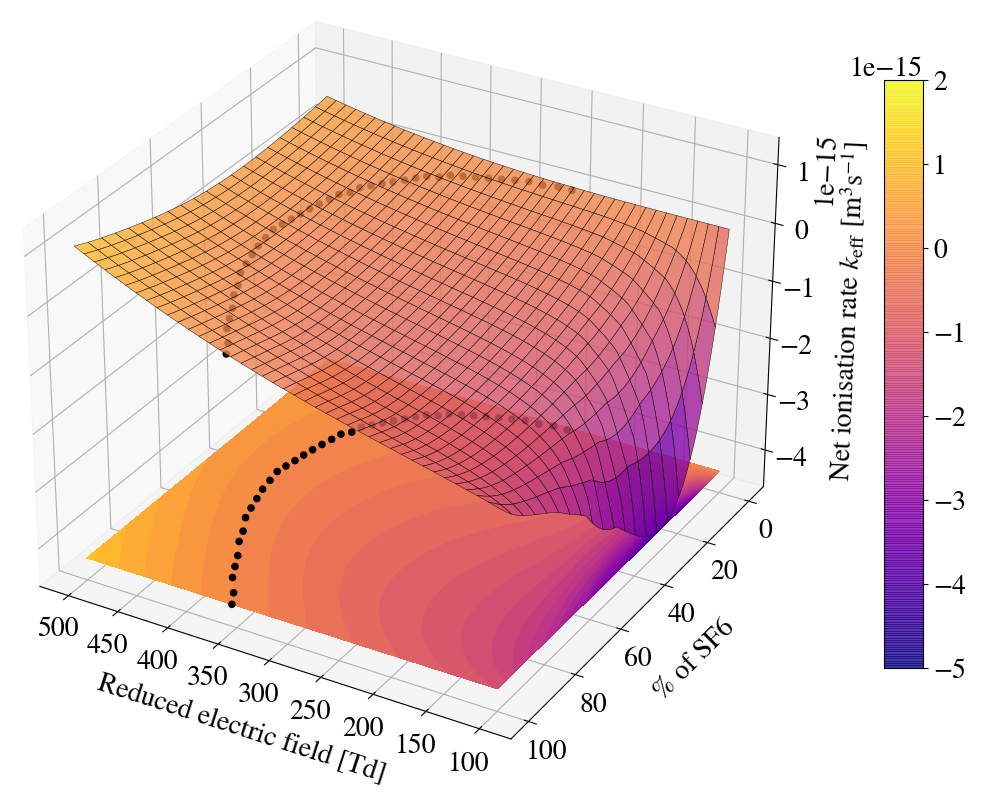}
    \caption{Surface of net effective ionisation rate, $k_{\mathrm{eff}}$, for mixtures of SF$_6$ and HFO-1234ze(E) as a function of $E/n_0$ and \% fraction of SF$_6$. Level curves projected on lower plane, with $k_{\mathrm{eff}}=0$ points marked in black,~\raisebox{0.5pt}{\tikz{\node[draw,scale=0.4,circle,fill=black!20!black](){};}}.}
    \label{fig:rate_field_scan_surface}
\end{figure}

\begin{figure}[H]
    \centering
    \includegraphics[width=0.9\linewidth]{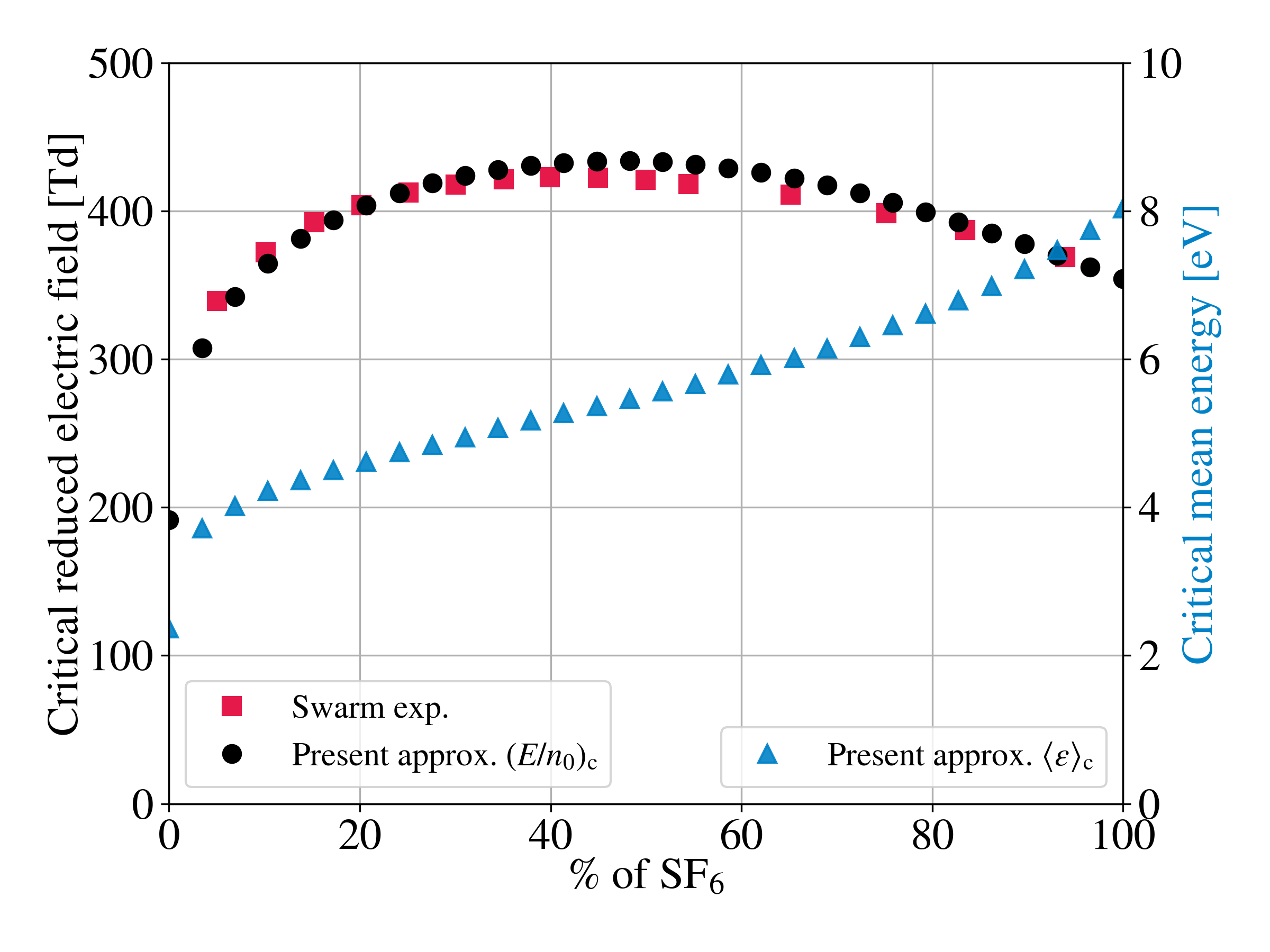}
    \caption{(Left-hand axis) Approximate critical reduced electric field where attachment transitions to ionisation for increasing fractions of SF$_6$ in mixture with  HFO-1234ze(E). Present approximation result shown in black markers,~\raisebox{0.5pt}{\tikz{\node[draw,scale=0.4,circle,fill=black!20!black](){};}}, swarm measurements of Egüz \textit{et al.}\cite{eguz_discussion_2022} shown in red markers,~\raisebox{0.5pt}{\tikz{\node[draw,scale=0.4,regular polygon, regular polygon sides=4,fill=red!10!red,line width=0](){};}}.  (Right-hand axis) Mean electron energy, $\langle \epsilon \rangle$, at critical electric field plotted in triangle markers,~\raisebox{0.5pt}{\tikz{\node[draw,scale=0.3,regular polygon, regular polygon sides=3,fill=blue!10!blue,rotate=0,line width=0](){};}}, to demonstrate monotonic nature}
    \label{fig:crit_field}
\end{figure}

The results in Figure~\ref{fig:rate_field}, and subsequent Figures~\ref{fig:rate_field_scan_surface}~and~\ref{fig:crit_field}, clearly demonstrate the well established non-monotonic behaviour of the critical field increasing with increasing SF$_6$ fraction but then reducing as the fraction of SF$_6$ approaches 100\%. The ability of these mixtures to yield a critical reduced field greater than either case of pure gas has been termed as a positive synergy in the insulating gas literature. This positive synergy effect has been well studied and discussed~\cite{eguz_discussion_2022,hosl_positive_2020,hunter1985} for insulating gases, and so being able to reproduce this important physics phenomenon from the present approximation formula, despite only using known data for pure gas cases, is an encouraging outcome of the present method. 

Despite the added complication of the non-monotonic positive synergy phenomenon our present method again provides a good approximation to the critical reduced electric field for these mixtures in the regime of competing ionising and attaching processes. We can see from Table~\ref{tab:sf6hfo} there is excellent quantitative agreement between the approximated value for the critical reduced electric field for mixtures, with relative errors on the order of 1-2\%.

\begin{table}[H]
    \centering
    \caption{Errors in critical reduced electric field approximation for various mixtures of SF$_6$ and HFO-1234ze(E).}
    \label{tab:sf6hfo}
    
    \begin{tabular}{cccc}
    \cline{2-3}
    \multicolumn{1}{c|}{}             & \multicolumn{2}{c|}{Critical $E/n_0$ {[}Td{]}}                                   &                                              \\ \cline{2-4} 
    \multicolumn{1}{c|}{}             & \multicolumn{1}{c|}{\textbf{Actual}} & \multicolumn{1}{c|}{\textbf{Approximate}} & \multicolumn{1}{c|}{\textbf{Relative error}} \\ \hline
    \multicolumn{1}{|c|}{25\% SF$_6$} & \multicolumn{1}{c|}{402}             & \multicolumn{1}{c|}{403}                  & \multicolumn{1}{c|}{0.2\%}                   \\ \hline
    \multicolumn{1}{|c|}{50\% SF$_6$} & \multicolumn{1}{c|}{426}             & \multicolumn{1}{c|}{434}                  & \multicolumn{1}{c|}{0.7\%}                   \\ \hline
    \multicolumn{1}{|c|}{75\% SF$_6$} & \multicolumn{1}{c|}{401}             & \multicolumn{1}{c|}{398}                  & \multicolumn{1}{c|}{1.9\%}                   \\ \hline
                                      &                                      &                                           &                                             
    \end{tabular}
\end{table}

\begin{figure}[H]
    \centering
    \includegraphics[width=1.1\linewidth]{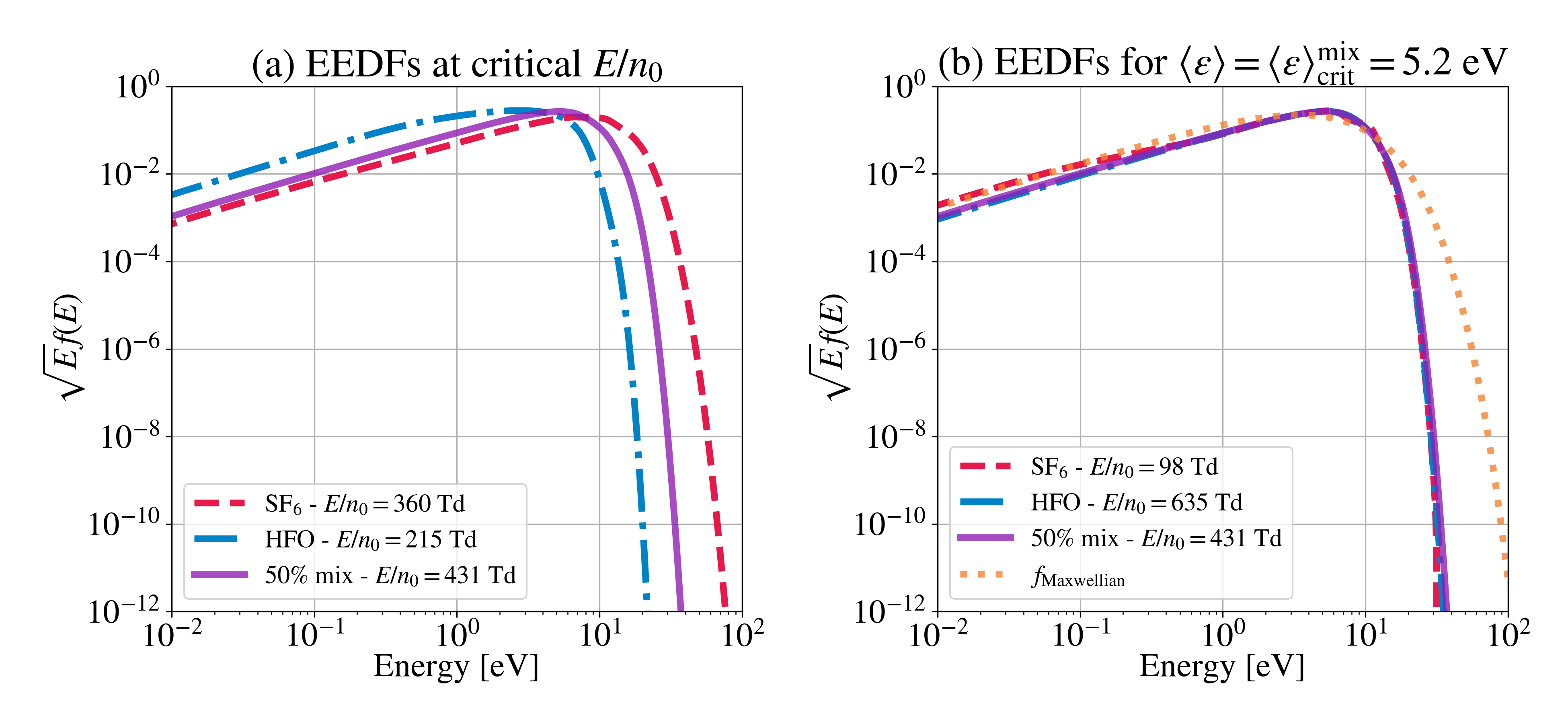}
    \caption{Comparison of EEDFs for SF$_6$, HFO-1234ze(E), and a mixture of 50\% SF$_6$ and 50\% HFO-1234ze(E). (Left) EEDFs at each critical $E/n_0$ value for the three background gas cases. (Right) EEDFs for each background gas case associated with a mean electron energy $\langle \epsilon \rangle=5.2$ eV corresponding to the mixture EEDF at the critical field threshold.}
    \label{fig:eedf2}
\end{figure}

Comparing the EEDFs, in Figure~\ref{fig:eedf2}(a), we can see quite different distributions for the case of pure SF$_6$, HFO-1234ze(E), and a mixture of 50\% SF$_6$ and 50\% HFO-1234ze(E). At low energies the mixture EEDF resembles that of SF$_6$, however beyond about 1 eV the mixture EEDF adopts qualitative features of the pure HFO-1234ze(E) distribution. To compare EEDFs corresponding to the same common electron mean energy, we refer to subfigure (b). Here, we see that when considering EEDFs yielding the same mean energy as the mixture critical point, $\langle \epsilon \rangle=5.2$ eV, the distributions are once again very similar. Once again, this outcome validates the use of the underlying common mean energy assumption discussed in Section~\ref{sec:theory}.

We now demonstrate an example of high-throughput data generation and understanding that the present semi-analytic method offers, compared to having to calculate the exact EEDF via the Boltzmann equation for each system configuration. Scanning over a range of mixture combinations $0\leq x \leq 1$ and a range of applied $E/n_0$ values allows a surface of the net effective ionisation rate, $k_{\mathrm{eff}}$, to be constructed over the domain of possible parameter space, as shown in Figure~\ref{fig:rate_field_scan_surface}. Considering the colour gradient changes projected to the base of the plot, one can clearly see the non-linear and non-monotonic features of the levels of $k_{\mathrm{eff,m}}$.  

For the possible mixture combinations in the case of $k_{\mathrm{eff,m}}\left(E/n_0\right)=0$, black marker points are placed to highlight the critical field curve for the mixtures of SF$_6$ and HFO-1234ze(E). These points can be projected onto a more convenient 2D plot, like that commonly presented in literature, in Figure~\ref{fig:crit_field}. For comparison the pulsed Townsend swarm measurements of Egüz\textit{ et al.}~\cite{eguz_synergism_2023} have been plotted. Further, the critical electron mean energy, $\langle \epsilon \rangle_c$, at a given critical electric field is plotted against the right-hand axis of Figure~\ref{fig:crit_field}.

In contrast to the non-monotonic nature of \ecrit, the critical electron mean energy, $\langle \epsilon \rangle_c$, plotted in Figure~\ref{fig:crit_field} demonstrates a monotonic increase with increasing fraction of SF$_6$. By plotting and comparing \ecrit ~and the equivalent critical $\langle \epsilon \rangle_c$ we highlight why indeed a mixture approximation of linear addition of constituent gas quantities with functional dependence on $E/n_0$ may fail. In contrast, the monotonic nature in $\langle \epsilon \rangle_c$ allows a linear combination of pure gas quantities with mean energy functional dependence to be used effectively. 

\section{Conclusion} \label{sec:conclusion}
In summary, we have presented the formulation and validation of a semi-analytic formula and method for approximating the critical reduced electric field for insulating gas mixtures. These findings demonstrate a rapid and accurate approximate approach to assessing the breakdown electric field properties of arbitrary insulating gas mixtures. Importantly, our results show immediate application and promise towards parameter space exploration and design studies of potential insulating gases to replace SF$_6$ in the high voltage electricity industry. While this study provides valuable insights into how industry standard electron transport data of individual pure gases can be combined to estimate mixture properties, further investigation is needed, for example, to develop improvements to the model to facilitate the use of experimentally measured swarm quantities as a function of $E/n_0$. While high-throughput approximation and prototyping can be done using the presented method, the necessity of having calculated pure gas swarm data as a function of mean energy is a limitation. Future studies will explore this by studying functional dependencies of electron mean energy, characteristic energy, and quantities measured directly in swarm experiments. By continuing to investigate these aspects, we can expand the set of quality tools available to scientists and engineers in the design and development of effective, safe, and climate neutral insulating gas technologies for high voltage electricity industries. 

\noindent\rule{\textwidth}{1pt}

\section*{References} \label{sec:refs}
\bibliography{references}

\begin{thebibliography}{10}

\bibitem{tian_research_2020}
Shuangshuang Tian, Xiaoxing Zhang, Yann Cressault, Juntai Hu, Bo~Wang, Song
  Xiao, Yi~Li, and Narjisse Kabbaj.
\newblock Research status of replacement gases for {SF6} in power industry.
\newblock {\em AIP Advances}, 10(5):050702, May 2020.

\bibitem{noauthor_kyoto_2005}
{United} {Nations} {Framework}~{Convention} on~{Climate}~{Change}.
\newblock A {Kyoto} {Protocol} to the {United} {Nations} {Framework}
  {Convention} on {Climate} {Change} {Kyoto}, 11 {December} 1997.
\newblock In {\em A {Kyoto} {Protocol} to the {United} {Nations} {Framework}
  {Convention} on {Climate} {Change} {Kyoto}, 11 {December} 1997}, volume 2303
  of {\em Treaty {Series}}, page 162. United Nations, 2005.

\bibitem{hu_declining_2023}
Lei Hu, Deborah Ottinger, Stephanie Bogle, Stephen~A. Montzka, Philip~L.
  DeCola, Ed~Dlugokencky, Arlyn Andrews, Kirk Thoning, Colm Sweeney, Geoff
  Dutton, Lauren Aepli, and Andrew Crotwell.
\newblock Declining, seasonal-varying emissions of sulfur hexafluoride from the
  {United} {States}.
\newblock {\em Atmospheric Chemistry and Physics}, 23(2):1437--1448, January
  2023.
\newblock Publisher: Copernicus GmbH.

\bibitem{simmonds_increasing_2020}
Peter~G. Simmonds, Matthew Rigby, Alistair~J. Manning, Sunyoung Park, Kieran~M.
  Stanley, Archie McCulloch, Stephan Henne, Francesco Graziosi, Michela Maione,
  Jgor Arduini, Stefan Reimann, Martin~K. Vollmer, Jens Mühle, Simon
  O'Doherty, Dickon Young, Paul~B. Krummel, Paul~J. Fraser, Ray~F. Weiss,
  Peter~K. Salameh, Christina~M. Harth, Mi-Kyung Park, Hyeri Park, Tim Arnold,
  Chris Rennick, L.~Paul Steele, Blagoj Mitrevski, Ray H.~J. Wang, and
  Ronald~G. Prinn.
\newblock The increasing atmospheric burden of the greenhouse gas sulfur
  hexafluoride ({SF}$_{\textrm{6}}$).
\newblock {\em Atmospheric Chemistry and Physics}, 20(12):7271--7290, June
  2020.
\newblock Publisher: Copernicus GmbH.

\bibitem{basu_improved_2023}
Devayan Basu, Juriy Pachin, Eda Egüz, and Christian Franck.
\newblock Improved {Estimation} of {Uniform} {AC} {Electric} {Breakdown}
  {Field} {Strength} of {HFO1234ze}({E}).
\newblock {\em IEEE Transactions on Dielectrics and Electrical Insulation},
  30(1):263--270, February 2023.
\newblock Conference Name: IEEE Transactions on Dielectrics and Electrical
  Insulation.

\bibitem{zeng_breakdown_2022}
Fuping Zeng, Bingquan Xie, Dazhi Su, Chen Li, Zhicheng Lei, Guoming Ma,
  Liangjun Dai, Long Li, and Ju~Tang.
\newblock Breakdown {Characteristics} of {Eco}-{Friendly} {Gas} {C5F10O}/{CO2}
  {Under} {Switching} {Impulse} in {Nonuniform} {Electric} {Field}.
\newblock {\em IEEE Transactions on Dielectrics and Electrical Insulation},
  29(3):866--873, June 2022.
\newblock Conference Name: IEEE Transactions on Dielectrics and Electrical
  Insulation.

\bibitem{eguz_synergism_2023}
E.~A. Egüz, J.~Pachin, H.~Vemulapalli, and C.~M. Franck.
\newblock Synergism in {SF6} mixtures with {C}={C}-{C} backbone compounds.
\newblock {\em Journal of Physics D: Applied Physics}, 56(39):395203, July
  2023.
\newblock Publisher: IOP Publishing.

\bibitem{eguz_discussion_2022}
E~A Egüz, J~Pachin, and C~M Franck.
\newblock Discussion on the mechanism leading to positive synergism in {SF}
  $_{\textrm{6}}$ mixtures with {HFO1234ze}({E}).
\newblock {\em Journal of Physics D: Applied Physics}, 55(31):315203, August
  2022.

\bibitem{zhang_numerical_2022}
Runming Zhang, Lijun Wang, Jie Liu, and Zhuoxi Lian.
\newblock Numerical simulation of breakdown properties and streamer development
  processes in {SF6}/{CO2} mixed gas.
\newblock {\em AIP Advances}, 12(1):015003, January 2022.

\bibitem{flynn_benchmarking_2021}
M.~Flynn, A.~Neuber, and J.~Stephens.
\newblock Benchmarking the calculation of electrically insulating properties of
  complex gas mixtures using a multi-term {Boltzmann} equation model.
\newblock {\em Journal of Physics D: Applied Physics}, 55(1):015201, October
  2021.
\newblock Publisher: IOP Publishing.

\bibitem{chachereau_electron_2016}
A~Chachereau, M~Rabie, and C~M Franck.
\newblock Electron swarm parameters of the hydrofluoroolefine {HFO1234ze}.
\newblock {\em Plasma Sources Science and Technology}, 25(4):045005, August
  2016.

\bibitem{blanc_recherches_1908}
A.~Blanc.
\newblock Recherches sur les mobilités des ions dans les gaz.
\newblock {\em Journal de Physique Théorique et Appliquée}, 7(1):825--839,
  1908.
\newblock Publisher: Société Française de Physique.

\bibitem{wieland_gasdurchschlagmechanismen_1973}
A~Wieland.
\newblock Gasdurchschlagmechanismen in elektronegativen {Gasen} ({SF6}) und in
  {Gasgemischen}.
\newblock {\em ETZ-A}, 94(7):370--373, 1973.

\bibitem{maric_parametrization_2005}
D.~Marić, M.~Radmilović-Radenović, and Z.~Lj. Petrović.
\newblock On parametrization and mixture laws for electron ionization
  coefficients.
\newblock {\em The European Physical Journal D - Atomic, Molecular, Optical and
  Plasma Physics}, 35(2):313--321, August 2005.

\bibitem{van_brunt_common_1987}
R.~J. Van~Brunt.
\newblock Common parametrizations of electron transport, collision cross
  section, and dielectric strength data for binary gas mixtures.
\newblock {\em Journal of Applied Physics}, 61(5):1773--1787, March 1987.

\bibitem{chiflikian_analog_1995}
R.~V. Chiflikian.
\newblock The analog of {Blanc}’s law for drift velocities of electrons in
  gas mixtures in weakly ionized plasma.
\newblock {\em Physics of Plasmas}, 2(10):3902--3909, October 1995.

\bibitem{garland_unified_2017}
N.~A. Garland, D.~G. Cocks, G.~J. Boyle, S.~Dujko, and R.~D. White.
\newblock Unified fluid model analysis and benchmark study for electron
  transport in gas and liquid analogs.
\newblock {\em Plasma Sources Science and Technology}, 26(7):075003, June 2017.
\newblock Publisher: IOP Publishing.

\bibitem{simonovic_electron_2019}
I.~Simonović, N.~A. Garland, D.~Bošnjaković, Z.~Lj Petrović, R.~D. White,
  and S.~Dujko.
\newblock Electron transport and negative streamers in liquid xenon.
\newblock {\em Plasma Sources Science and Technology}, 28(1):015006, January
  2019.
\newblock Publisher: IOP Publishing.

\bibitem{boyle_boltzmanns_2023}
G.~J. Boyle, P.~W. Stokes, R.~E. Robson, and R.~D. White.
\newblock Boltzmann’s equation at 150: {Traditional} and modern solution
  techniques for charged particles in neutral gases.
\newblock {\em The Journal of Chemical Physics}, 159(2):024306, July 2023.

\bibitem{garland_electron_2018}
N.~A. Garland, I.~Simonović, G.~J. Boyle, D.~G. Cocks, S.~Dujko, and R.~D.
  White.
\newblock Electron swarm and streamer transport across the gas–liquid
  interface: a comparative fluid model study.
\newblock {\em Plasma Sources Science and Technology}, 27(10):105004, October
  2018.
\newblock Publisher: IOP Publishing.

\bibitem{garland_approximating_2018}
N.~A. Garland, G.~J. Boyle, D.~G. Cocks, and R.~D. White.
\newblock Approximating the nonlinear density dependence of electron transport
  coefficients and scattering rates across the gas–liquid interface.
\newblock {\em Plasma Sources Science and Technology}, 27(2):024002, February
  2018.
\newblock Publisher: IOP Publishing.

\bibitem{hagelaar_solving_2005}
G.~J.~M. Hagelaar and L.~C. Pitchford.
\newblock Solving the {Boltzmann} equation to obtain electron transport
  coefficients and rate coefficients for fluid models.
\newblock {\em Plasma Sources Science and Technology}, 14(4):722--733, October
  2005.
\newblock Publisher: IOP Publishing.

\bibitem{rabie_methes_2016}
M.~Rabie and C.~M. Franck.
\newblock {METHES}: {A} {Monte} {Carlo} collision code for the simulation of
  electron transport in low temperature plasmas.
\newblock {\em Computer Physics Communications}, 203:268--277, June 2016.

\bibitem{stephens_multi-term_2018}
J.~Stephens.
\newblock A multi-term {Boltzmann} equation benchmark of electron-argon
  cross-sections for use in low temperature plasma models.
\newblock {\em Journal of Physics D: Applied Physics}, 51(12):125203, March
  2018.
\newblock Publisher: IOP Publishing.

\bibitem{park2023thunderboltz}
Ryan Park, Brett~S. Scheiner, and Mark~C. Zammit.
\newblock Thunderboltz: An open-source {DSMC}-based {Boltzmann} solver for
  plasma transport, chemical kinetics, and {0D} plasma modeling.
\newblock {\em arXiv preprint}, 2310.07913, 2023.

\bibitem{white_electron_2018}
R.~D. White, D.~Cocks, G.~Boyle, M.~Casey, N.~Garland, D.~Konovalov,
  B.~Philippa, P.~Stokes, J.~de Urquijo, O.~González-Magaña, R.~P. McEachran,
  S.~J. Buckman, M.~J. Brunger, G.~Garcia, S.~Dujko, and Z.~Lj Petrovic.
\newblock Electron transport in biomolecular gaseous and liquid systems:
  theory, experiment and self-consistent cross-sections.
\newblock {\em Plasma Sources Science and Technology}, 27(5):053001, May 2018.
\newblock Publisher: IOP Publishing.

\bibitem{boyle_multi-term_2017}
G.~J. Boyle, W.~J. Tattersall, D.~G. Cocks, R.~P. McEachran, and R.~D. White.
\newblock A multi-term solution of the space–time {Boltzmann} equation for
  electrons in gases and liquids.
\newblock {\em Plasma Sources Science and Technology}, 26(2):024007, January
  2017.
\newblock Publisher: IOP Publishing.

\bibitem{pitchfordlxcat}
Leanne~C. Pitchford, Luis~L. Alves, Klaus Bartschat, Stephen~F. Biagi,
  Marie-Claude Bordage, Igor Bray, Chris~E. Brion, Michael~J. Brunger, Laurence
  Campbell, Alise Chachereau, Bhaskar Chaudhury, Loucas~G. Christophorou, Emile
  Carbone, Nikolay~A. Dyatko, Christian~M. Franck, Dmitry~V. Fursa, Reetesh~K.
  Gangwar, Vasco Guerra, Pascal Haefliger, Gerjan J.~M. Hagelaar, Andreas
  Hoesl, Yukikazu Itikawa, Igor~V. Kochetov, Robert~P. McEachran, W.~Lowell
  Morgan, Anatoly~P. Napartovich, Vincent Puech, Mohamed Rabie, Lalita Sharma,
  Rajesh Srivastava, Allan~D. Stauffer, Jonathan Tennyson, Jaime de~Urquijo,
  Jan van Dijk, Larry~A. Viehland, Mark~C. Zammit, Oleg Zatsarinny, and Sergey
  Pancheshnyi.
\newblock Lxcat: an open-access, web-based platform for data needed for
  modeling low temperature plasmas.
\newblock {\em Plasma Processes and Polymers}, 14(1-2):1600098, 2017.

\bibitem{chachereau_characterization_2017}
Alise Chachereau and Christian Franck.
\newblock Characterization of {HFO1234ze} mixtures with {N2} and {CO2} for use
  as gaseous electrical insulation media.
\newblock page 6 p., 2017.
\newblock Artwork Size: 6 p. Medium: application/pdf Publisher: ETH Zurich.

\bibitem{li_partial_2023}
Yi~Li, Yifan Wang, Song Xiao, Zhen Li, Nian Tang, Yongyan Zhou, Li~Li, Yifan
  Zhang, Ju~Tang, and Xiaoxing Zhang.
\newblock Partial discharge induced decomposition and by-products generation
  properties of {HFO}-1234ze({E})/{CO2}: a new eco-friendly gas insulating
  medium.
\newblock {\em Journal of Physics D: Applied Physics}, 56(16):165203, March
  2023.
\newblock Publisher: IOP Publishing.

\bibitem{hosl_positive_2020}
Andreas Hosl, Juriy Pachin, Eda Eguz, Alise Chachereau, and Christian~M.
  Franck.
\newblock Positive synergy of {SF} $_{\textrm{6}}$ and {HFO1234ze}({E}).
\newblock {\em IEEE Transactions on Dielectrics and Electrical Insulation},
  27(1):322--324, February 2020.

\bibitem{muccignat2023iterative}
Dale~L Muccignat, Gregory~G Boyle, Nathan~A Garland, Peter~W Stokes, and
  Ronald~D White.
\newblock An iterative deep learning procedure for determining electron
  scattering cross-sections from transport coefficients.
\newblock {\em arXiv preprint}, 2311.13170, 2023.

\bibitem{stokes_determining_2020}
P.~W. Stokes, D.~G. Cocks, M.~J. Brunger, and R.~D. White.
\newblock Determining cross sections from transport coefficients using deep
  neural networks.
\newblock {\em Plasma Sources Science and Technology}, 29(5):055009, May 2020.
\newblock Publisher: IOP Publishing.

\bibitem{stokes_self-consistent_2020}
P.~W. Stokes, M.~J.~E. Casey, D.~G. Cocks, J.~de Urquijo, G.~García, M.~J.
  Brunger, and R.~D. White.
\newblock Self-consistent electron–{THF} cross sections derived using
  data-driven swarm analysis with a neural network model.
\newblock {\em Plasma Sources Science and Technology}, 29(10):105008, October
  2020.
\newblock Publisher: IOP Publishing.

\bibitem{hunter1985}
S.~R. Hunter and L.~G. Christophorou.
\newblock {Pressure‐dependent electron attachment and breakdown strengths of
  unary gases and synergism of binary gas mixtures: A relationship}.
\newblock {\em Journal of Applied Physics}, 57(9):4377--4385, 05 1985.

\end{thebibliography}
\bibliographystyle{unsrt}

\end{document}